\documentclass[runinaddress,showpacs,twocolumn,prb]{revtex4}
\usepackage{amsmath}
\usepackage{amssymb}
\usepackage{graphicx}
\usepackage{bm}

\begin{document}

\title{Spin-Current-Induced Charge Accumulation and Electric Current in
Semiconductor Nanostructures with Rashba Spin-Orbit Coupling}
\author{Jian Li and Shun-Qing Shen}
\affiliation{Department of Physics, and Center for Theoretical and Computational Physics,
The University of Hong Kong, Pokfulam Road, Hong Kong, China}
\date{November 29, 2006}

\begin{abstract}
We demonstrate that the flow of a longitudinal spin current with different
spin polarization will induce different patterns of charge accumulation in a
two-terminal strip, or electric current distribution in a four-terminal
Hall-bar structure, of two-dimensional electron gas with Rashba spin-orbit
coupling (RSOC). For an in-plane polarized spin current, charges will
accumulate either by the two lateral edges or around the center of the strip
structure while, for an out-of-plain polarized spin current, charge
densities will show opposite signs by the two lateral edges leading to a
Hall voltage. Our calculation offers a new route to experimentally detect or
differentiate pure spin currents with various spin polarization.
\end{abstract}

\maketitle

Semiconductor spintronics has achieved remarkable success in the
past decade and is still progressing rapidly. And spin-orbit
science and engineering, which allow for electrical manipulation
of spin polarization and spin currents in nonmagnetic
semiconductors, is one of the key steps to implement spintronic
devices.\cite{Prinz98Science} Utilizing spin-orbit coupling,
various schemes were proposed to generate pure spin
currents,\cite{Dyakonov71} while to detect pure spin currents
remains a challenge from either experimental or theoretical
aspect. The detection of spin currents often involves spin
accumulation or electrical effects, despite quantum interferences
by optical means also reported.\cite{Hubner03prl} For example,
spin accumulation induced by spin current near the boundaries has
been detected in both n- and p-doped semiconductor systems
experimentally,\cite{Kato04Science} and Hall voltage resulted from
the reciprocal extrinsic spin Hall effect was observed in
diffusive metallic conductors.\cite{Valenzuela06,Saitoh06apl}
Recently it was reported that an optically injected spin current
flowing through a Hall-bar semiconductor can generate either
in-ward or out-ward electric currents while the Hall voltage
remains zero.\cite{Cui06} This observation renders a manifestation
of the tensor-like nature of the spin current, which means that
both the spin polarization and the direction of motion are
decisive factors in producing physically observable effects.

\begin{figure}[tbph]
\centering \includegraphics[width=0.45\textwidth]{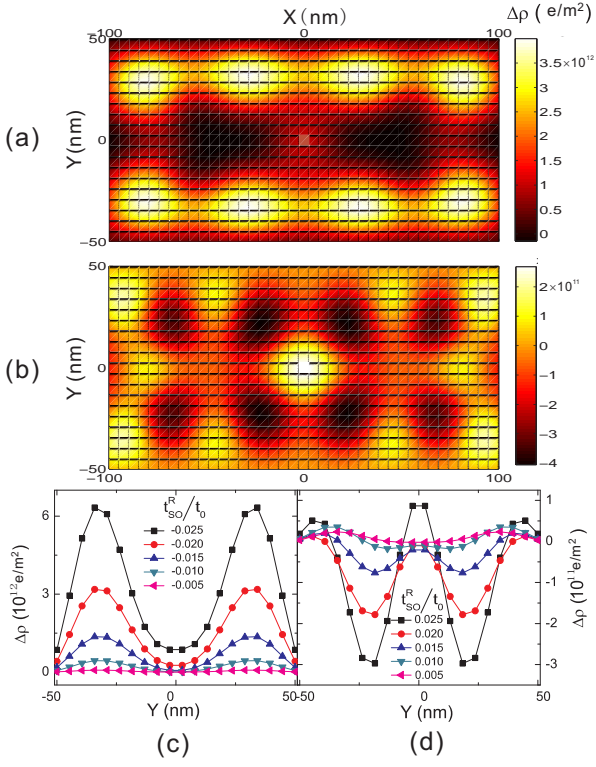}\newline
\caption{Spin-current-induced charge distributions in a 200nm$\times $100nm
strip, where the spin current is polarized along $\hat{y}$ and the
distributions are displayed after the spin-independent background
distribution (when $\protect\alpha =0$) has been subtracted from. (a) the
image of charge distribution when $\protect\alpha =-6.1\times 10^{-12}$eVm.
(b) the image of charge distribution when $\protect\alpha =6.1\times
10^{-12} $eVm. (c) and (d) the averaged (in terms of the longitudinal
dimension) charge distributions along the narrow side of the sample,
comparing different cases with $\protect\alpha $ having the same sign
(negative for plot (c) and positive for plot (d)) but increasing magnitudes.
}
\label{fig::cdy}
\end{figure}

The electrical detection of spin currents has reduced complexities
in practice and thus is potentially more
applicable.\cite{Valenzuela06, Saitoh06apl, Cui06} Whereas a
fundamental problem of theoretically studying the electrical
effects resulted from spin currents, is the difficulties to
incorporate the concept of spin current into a theoretical
formalism, for reasons like the ambiguous definition of spin
currents under certain circumstances.\cite{Shi06prl} In this
Letter we investigate such effects in a mesoscopic system of
spin-orbit-coupled two-dimensional gas (2DEG) with ideal leads
which connect several special electron reservoirs. Generation of
the spin current is simulated phenomenologically by introducing
spin-dependent chemical potentials within each electron reservoir.
These chemical potentials are tuned separately for each spin
component to produce independent potential gradients, so that
electrons with opposite spins are driven to move in opposite
directions through the spin-conserved leads and the
spin-orbit-coupled central region. It must be emphasized that the
aim of this phenomenological simulation is to capture only the
general and essential features of the spin current flow, without
bothering about the details of the method in its generation and
injection, and this is justified as long as we keep our focus on
the electrical effects that are led by or intimately related to
the \textit{circulation} of the spin current.

We demonstrate spin-current-induced electrical effects by showing charge
accumulation induced in a strip structure and the electric current
distribution in a corresponding Hall-bar structure. The Landauer-B\"{u}%
ttiker-Keldysh formalism is used in our calculations, which is a
quantum-in-nature approach and has extensive application.\cite{Sheng05prl}
The total Hamiltonian of the central region is $H_{C}=H_{0}+H_{SO}$ , where $%
H_{0}=\frac{\hbar ^{2}k^{2}}{2m^{\ast }}+V_{0}$ is the kinetic energy plus
the hard-wall confining potential $V_{0}$, $m^{\ast }$ is the effective
electron mass, and $H_{SO}=-\alpha (\bm{k}\times \bm{\sigma})\cdot \hat{z}$,
with $\alpha $ the strength of Rashba spin-orbit coupling (RSOC), $\bm{k}$
the wave vector, $\bm{\sigma}$ the vector composed of Pauli's matrices, and $%
\hat{z}$ the unit vector perpendicular to the plane of the 2DEG. After
discretizing $H_{C}$ with the tight-binding approximation and transforming
it into the spin bases which are the eigenstates (denoted by $\chi _{\mu }$
or $\chi _{\nu }$ where $\mu ,\nu =\uparrow $ or $\downarrow $) of $\hat{r}%
\cdot \bm{\sigma}$, where $\hat{r}$ is the orientation of spin polarization
under consideration, we have
\begin{equation}
H_{C}=\sum\limits_{<\mathbf{ij}>}\sum\limits_{\mu ,\nu }t_{\mathbf{ij}}^{\mu
\nu }c_{\mathbf{i}\mu }^{\dagger }c_{\mathbf{j}\nu }  \label{eq::Hc}
\end{equation}%
where $c^{\dag }$ and $c$ are the creation and annihilation operators of
electrons at sites $\mathbf{i}(\mathbf{j})$ with spin $\mu (\nu )$ and $%
<\cdots >$ means the pairs of nearest neighboring (n.n.) sites, and
\begin{equation*}
t_{\mathbf{ij}}^{\mu \nu }=\left\{
\begin{array}{ll}
(u^{\dagger })_{\mu \alpha }\bigl(-t_{0}I\mp it_{SO}^{R}\sigma _{y}\bigr)%
_{\alpha \beta }u_{\beta \nu }, & \quad \mathbf{i}=\mathbf{j}\pm \vec{\delta}%
_{x} \\
(u^{\dagger })_{\mu \alpha }\bigl(-t_{0}I\pm it_{SO}^{R}\sigma _{x}\bigr)%
_{\alpha \beta }u_{\beta \nu }, & \quad \mathbf{i}=\mathbf{j}\pm \vec{\delta}%
_{y}%
\end{array}%
\right.
\end{equation*}%
where $\vec{\delta}_{x(y)}$ is the unit vector displacement between two n.n.
sites in $x(y)$ direction, $t_{0}=\hbar ^{2}/2m^{\ast }a^{2}$ is the overlap
integral of two n.n. sites with $a$ the average spacing between two n.n.
sites, $t_{SO}^{R}=\alpha /2a$, $I$ is the identity matrix, and $u=(\chi
_{\uparrow },\chi _{\downarrow })$ is the unitary matrix which rotates $%
\sigma _{z}$ to $\hat{r}\cdot \bm{\sigma}$. To count in the effect
of the semi-infinite ideal leads, self-energy terms are
introduced, $\Sigma ^{r}=\sum \Sigma _{p}^{r}$, with the specific
term due to lead $p$ in the $\mu$ spin diagonal block is
\begin{equation}
\Sigma _{p,\mu }^{r}(\mathbf{i},\mathbf{j};E)=-t_{0}\sum\limits_{m}\phi
_{m}(p_{\mathbf{i}})e^{ik_{m}(E)a}\phi _{m}(p_{\mathbf{j}})  \label{eq::SEp}
\end{equation}%
where $\phi _{m}(p_{i})$ is the $m^{th}$ eigenfunction in the transverse
dimension at site $p_{\mathbf{i}}$ in lead $p$ which is adjacent to site $%
\mathbf{i}$ in the central region, and $k_{m}$ is the wave vector
along the semi-infinite lead. Here we assume that there is no
spin-orbit coupling in the leads. This not only guarantees that
the spin currents under our investigation is well-defined from
experimental aspect, but also is justified because it turns out
that the interface mismatch between the leads and the central part
actually contributes little to the patterns we observed. The
retarded Green's function $G^{r}(E)=(E-H_{C}-\Sigma ^{r})^{-1}$,
and the lesser Green's function is given by the Keldysh equation
$G^{<}=G^{r}\Sigma ^{<}G^{a}$, with $\Sigma ^{<}(E)=-\sum_{p,\mu
}f(E-\epsilon _{p}^{\mu })\bigl(\Sigma _{p,\mu }^{r}(E)-\Sigma
_{p,\mu }^{a}(E)\bigr)$, where $\epsilon _{p}^{\mu }$ is the
spin-dependent chemical potential for electrons of spin $\mu $ in
the lead $p$, and $f(E-\epsilon _{p}^{\mu })$ is the Fermi-Dirac
distribution function. Expressed in terms of the lesser Green's
function, when a steady state is achieved, the non-equilibrium
charge density at site $\mathbf{i}$ is
\begin{equation}
\rho _{c}(\mathbf{i})=-\frac{ie}{2\pi }\sum\limits_{\mu }\int_{-\infty
}^{\infty }dE\,G^{<}(\mathbf{i},\mu ;\mathbf{i},\mu ;E)  \label{eq::cd}
\end{equation}%
and the electric bond current from site $\mathbf{i}$ to site $\mathbf{j}$ is
\begin{eqnarray}
\displaystyle j_{\mathbf{ij}}^{c} &=&-\frac{e}{2\pi }\sum\limits_{\mu \nu
}\int_{-\infty }^{\infty }dE\,\Bigl[t_{\mathbf{ji}}^{\nu \mu }G^{<}(\mathbf{i%
},\mu ;\mathbf{j},\nu ;E)  \label{eq::ccd} \\
&&\;\displaystyle-t_{\mathbf{ij}}^{\nu \mu }G^{<}(\mathbf{j},\nu ;\mathbf{i}%
,\mu ;E)\Bigr]  \notag
\end{eqnarray}

Our present study focuses on electrical effects resulted from spin current
with spin polarization orientated in plane (referred to as $\hat{y}$
henceforth) which is perpendicular to the direction of motion (referred to
as $\hat{x}$ henceforth) of electron spin. When a spin current with $\hat{y}$%
-spin-polarization is flowing through a strip, depending upon the sign of
the RSOC $\alpha $, which stands for whether the velocity, the spin
polarization and the gradient of potential for the RSOC satisfy the
left-hand or right-hand chirality, the electric charges will accumulate
either near the lateral edges or around the middle of stripe, as shown in
Fig.\ref{fig::cdy}. In this figure, two equal-magnitude $+\hat{y}$/$-\hat{y}$%
-spin-polarized current are driven in the direction $+\hat{x}$/$-\hat{x}$ by
the spin-dependent chemical potentials $\epsilon _{\pm x}^{\mu }=\pm \,%
\mbox{sign}(\mu )0.1t_{0}$ through a $40\times 20$ lattice with the Fermi
Energy $E_{f}=0.1t_{0}$, which is measured from the bottom of the conduction
band and is small enough to ensure the parabolic energy-momentum dispersion
in the tight-binding approximation, and $\alpha =-6.1\times 10^{-12}$eVm or $%
t_{SO}^{R}=-0.02t_{0}$ in Fig.\ref{fig::cdy}(a), and $\alpha =+6.1\times
10^{-12}$eVm or $t_{SO}^{R}=+0.02t_{0}$ in Fig.\ref{fig::cdy}(b),
respectively. It is easy to identify that the charges of carriers accumulate
by the two lateral edges in the former case and at the middle part in the
latter one. This is further manifested by a comparison of Fig.\ref{fig::cdy}%
(c) and Fig.\ref{fig::cdy}(d), where the averaged charge densities with $\pm
\alpha $ are plotted. For a negative $\alpha $, increased magnitude will
lead to increased accumulation by the edges, while for a positive $\alpha $,
the same trend happens at the middle. It should be noted that since we do
not consider any dissipation mechanisms inside our samples, these
accumulations should be interpreted as an effect purely due to quantum
coherent transport.

\begin{figure}[tbph]
\centering \includegraphics[width=0.4\textwidth]{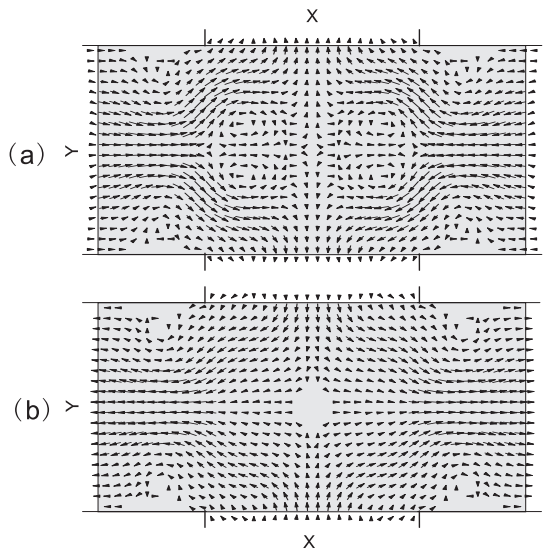}\newline
\caption{Charge current distributions after two additional leads are
connected to the lateral edges of the strip in Fig. 1. The direction and
magnitude of local current density are indicated by the orientation and
length of arrows, respectively. The transverse electric currents are both
flowing outward in plot (a), where $\protect\alpha =-6.1\times 10^{-12}$eVm
and the ratio of the total induced electric current to the circulating spin
current $I_{c}/I_{s}=2.57\times 10^{-2}\frac{e}{\hbar /2}$, or both flowing
inward in plot (b), where $\protect\alpha =6.1\times 10^{-12}$eVm and $%
I_{c}/I_{s}=1.14\times 10^{-2}\frac{e}{\hbar /2}$. }
\label{fig::ccdy}
\end{figure}

To show physical consequences of charge accumulation in a strip structure,
we calculated electric current distributions in a Hall-bar structure where
two additional leads are symmetrically attached to the two lateral sides of
the original strip as illustrated in Fig.\ref{fig::ccdy}. The current
distribution for either sign of $\alpha $ in this structure has remarkable
consistence with the charge accumulation in the corresponding strip system
in terms of that the additional leads act just as the pathways for the
accumulated charges to flow through. Specifically, when $\alpha <0$, charges
accumulated by the lateral edges will flow outward through the two
transverse leads and at the same time the net charges will be drawn though
the two longitudinal leads, while when $\alpha >0$, the charges tend to be
drawn inward through two transverse leads and flow out through two
longitudinal ones, which accounts for the accumulation of charges around the
center of the strip-shaped sample. Quantitatively, the induced electric
current in each transverse lead sums up to be typically two orders less than
the magnitude of the total circulating spin current. These observations are
also fully consistent with the electric current patterns in reference\cite%
{Li06apl} where the linear response approximation are adopted to produce the
results.

\begin{figure}[tbph]
\centering \includegraphics[width=0.45\textwidth]{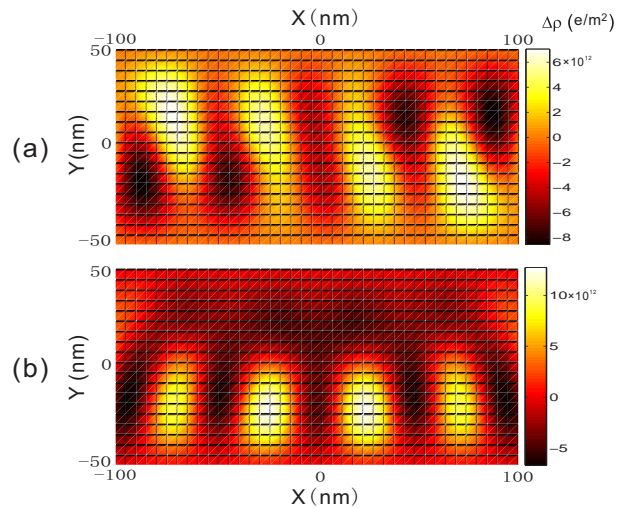}\newline
\caption{Spin-current-induced charge distributions in a 200nm$\times $100nm
strip, where the spin current is polarized along $\hat{x}$ in (a) and along $%
\hat{z}$ in (b). $\protect\alpha =6.1\times 10^{-12}$eVm and the
distributions are displayed after the spin-independent background
distribution has been subtracted from.}
\label{fig::cdxz}
\end{figure}

So far we have mainly discussed the charge density and the
electric current distribution induced by a spin current with
in-plane spin polarization (along $\hat{y}$). Yet a tensor as a
spin current is in essence, it is also worthwhile to investigate
the spin current with different configurations of spin
polarization and velocity. Without losing the generality, we
concentrate on three cases of spin along $\hat{x}$, $\hat{y}$ and
$\hat{z}$, respectively. From the study of these special cases,
the generic result for a spin current with arbitrary spin
polarization can be derived, and also
maximum symmetries can be observed therein. Since one of these cases with $%
\hat{y}$-spin-polarization has been presented already, we show the charge
distributions induced in other two cases in Fig.\ref{fig::cdxz}. Compared
with the induced charge distribution shown in Fig.\ref{fig::cdy}(b), which
is calculated with all the same parameters except for the spin polarization
of the spin current, the induced charge distributions in Fig.\ref{fig::cdxz}%
(a) and (b) show clear differences in terms of the spacial symmetries they
possess, that is the $C_{2v}$ symmetry in the case when spin polarized along
$\hat{y}$, the $C_{2}$ symmetry when spin polarized along $\hat{x}$, and the
$C_{s}$ symmetry when spin polarized along $\hat{z}$.\cite{Hamermesh} This
is because the underlying system is invariant under each operation of the $%
C_{2v}$ space group combined with an appropriate unitary transformation in
the spin space, while in order that the circulating spin current is also
invariant under that combination of transformations, the valid space-group
operations will be limited into a subgroup of $C_{2v}$ ($C_{2v}$ itself in
the $\hat{y}$-spin-polarization case, and $C_{2}$ or $C_{s}$ in the $\hat{x}$%
- or $\hat{z}$-spin-polarization case respectively), accordingly the charge
distributions induced by different spin currents will exhibit different
symmetry properties. Also it is easy to infer that the electric current
distributions or any other electric effect induced by a corresponding spin
current will display the same spacial symmetry as long as the structural
symmetry of the system is preserved, which is actually a general yet
rigorous limitation and has been justified in all our calculations that are
shown or not shown here.

Moreover, the relevance of the spin polarization of spin current to the
induced charge accumulation also lies in the differences between the average
distributions along the transverse of the strip resulted from spin currents
which are polarized in different directions. In particular, Fig.\ref%
{fig::cdy}(c) and (d) show that the charges of carriers tend to accumulate,
depending on the sign of $\alpha $, either by both of the two lateral edges
or around the center of the sample when the spin current is polarized along $%
\hat{y}$. When the spin current is polarized along $\hat{z}$, however, it
can be observed from Fig.\ref{fig::cdxz}(b) that the charge accumulation on
average will have opposite signs near the two lateral edges, which is
closely related to as well as consistent with the reciprocal version of the
spin Hall effect,\cite{Shi06prl,Hankiewicz05prb} and was reported to be
observed in platinum wires.\cite{Saitoh06apl} In another case that the spin
current with $\hat{x}$-spin-polarization is considered, the pattern of the
charge accumulation (shown in Fig.\ref{fig::cdxz}(a)) is more complex in the
sense that there are reversed accumulating trends at the two sides of the
transverse plane passing through the symmetric center, which is determined
by the $C_{2}$ symmetry of this system mentioned above. While after average
has been taken over the longitudinal dimension, it can be reasonably
expected that the opposite accumulations of two halves will tend to cancel
each other and produce a weakened net effect with symmetrical distribution
about the transversal center. In contrast to that in the $\hat{y}$%
-spin-polarized case, in either case with the $\hat{x}$- or $\hat{z}$%
-spin-polarization, the averaged distribution along the traverse does not
change under the reversal of the sign of $\alpha $. Besides, we notice that
the electric effects induced by different spin currents are actually
contributed by different ranks of power with respect to $\alpha $, which is
also manifested when the sign of $\alpha $ is reversed. Specifically, when
the spin current is polarized along $\hat{x}$ or $\hat{y}$, it is mainly the
linear $\alpha $ that is responsible for the induced electric effects, while
in the $\hat{z}$-spin-polarized case, it is the second rank, i.e. $\alpha
^{2}$, playing the role.

We conclude this letter by evaluating the accessibility of an
experimental observation to the electrical patterns investigated
here. From the data plotted in Fig.\ref{fig::cdy}(c), we estimate
roughly the electrostatic potential difference $\Delta V$ to be of
order $0.1$mV between either of the peaks and the valley of the
charge distribution as long as $\alpha >3\times 10^{-12}$eVm. And
in a realistic semiconductor quantum well with the Fermi energy
typically being tens of meV, the electronic potential difference
will increase to at least several meV or tens of Kelvins subject
to increased RSOC strength. This implies that the temperature
requirement for observing these patterns can be well satisfied
within current experimental conditions. Regarding these features
as well as the sample size, we propose the use of Kelvin probe
force microscopy\cite{KPFM} in observing the charge accumulation
predicted here. On the other hand, we point out that the present
work may also account for a recent experimental observation of the
spin-current-induced electric currents,\cite{Cui06} which possess
the key features exhibited in Fig.\ref{fig::cdy} and
Fig.\ref{fig::ccdy}, that is, \textit{there is no Hall voltage
while the electric currents circulate through
}$x$\textit{-channels to }$y$\textit{-channels}. And
quantitatively the experiment is consistent with the calculated
ratio of the induced charge current to the spin current captioned
in Fig.\ref{fig::cdy}. In short, the spin current with in-plane
spin polarization may produce measurable charge accumulations or
electric currents with a novel behavior, and the underlying
effects may open a promising way to the electrical detection of
spin currents.

The author would like to thank Xiao-Dong Cui and Fu-Chun Zhang for helpful
discussions. This work was supported by the Research Grant Council of Hong
Kong under Grant No.: HKU 0742/06P.


\begin{thebibliography}{99}
\bibitem{Prinz98Science} G. A. Prinz, Science 282, 1660 (1998); S. A. Wolf, D. D. Awschalom, R. A. Buhrman, J. M. Daughton, S. von Molnar, M. L. Roukes, A. Y. Chtchelkanova, and D. M. Treger, Science 294, 1488 (2001).

\bibitem{Dyakonov71} M. I. D'yakonov and V. I. Perel, JETP Lett. 13, 467
(1971); J. E. Hirsch, Phys. Rev. Lett. 83, 1834 (1999); J. Sinova,
D. Culcer, Q. Niu, N. A. Sinitsyn, T. Jungwirth, and A. H.
MacDonald, Phys. Rev. Lett. 92, 126603 (2004); S. Murakami, N.
Nagaosa, and S. C. Zhang, Science 301, 1348 (2003); S.-Q. Shen, M.
Ma, X. C. Xie, and F. C. Zhang, Phys. Rev. Lett. 92, 256603
(2004); R. D. R. Bhat and J. E. Sipe, Phys. Rev. Lett. 85, 5432
(2000); B. Wang, J. Peng, D. Y. Xing, and J. Wang, Phys. Rev.
Lett. 95, 086608 (2005); A. G. Mal'shukov, C. S. Tang, C. S. Chu,
and K. A. Chao, Phys. Rev. B 68, 233307 (2003)

\bibitem{Hubner03prl} J. Hubner, W. W. Ruhle, M. Klude, D. Hommel, R. D. R. Bhat, J. E. Sipe, and H. M. van Driel, Phys. Rev. Lett. 90, 216601
(2003); M. J. Stevens, A. L. Smirl, R. D. R. Bhat, A. Najmaie, J.
E. Sipe, and H. M. van Driel, Phys. Rev. Lett. 90, 136603 (2003);
H. Zhao, E. J. Loren, H. M. van Driel, and A. L. Smirl, Phys. Rev.
Lett. 96, 246601 (2006).

\bibitem{Kato04Science} Y. K. Kato, R. C. Myers, A. C. Gossard and D. D. Awschalom, Science 306, 1910
(2004); J. Wunderlich \textit{et al.}, Phys. Rev. Lett. 94, 047204
(2005); V. Sih, R. C. Myers, Y. K. Kato, W. H. Lau, A. C. Gossard,
and D. D. Awschalom, Nature Physics 1, 31 (2005).

\bibitem{Valenzuela06} S. O. Valenzuela and M Tinkham, Nature 442, 176
(2006).

\bibitem{Saitoh06apl} E. Saitoh, M. Ueda, H. Miyajima, and G. Tatara, Appl. Phys. Lett. 88,
182509 (2006); T. Kimura, Y. Otani, T. Sato, S. Takahashi, and S.
Maekawa, Phys. Rev. Lett. 98, 156601 (2007).

\bibitem{Cui06} X. D. Cui, S. Q. Shen, J. Li, Y. Ji, W. Ge, and F. C. Zhang, Appl. Phys. Lett. 90, 242115 (2007).

\bibitem{Shi06prl} J. R. Shi, P. Zhang, D. Xiao, and Q. Niu, Phys. Rev. Lett. 96, 076604
(2006).

\bibitem{Sheng05prl} L. Sheng, D. N. Sheng, and C. S. Ting, Phys. Rev. Lett. 94, 016602
(2005); B. K. Nikolic, S. Souma, L. P. Zarbo, and J. Sinova, Phys.
Rev. Lett. 95, 046601 (2005); E. M. Hankiewicz, L. W. Molenkamp,
T. Jungwirth, and J. Sinova, Phys. Rev. B 70, 241301(R) (2004); J.
Li, L. Hu, and S. Q. Shen, Phys. Rev. B 71, 241305(R) (2005).

\bibitem{Li06apl} J. Li, X. Dai, S. Q. Shen, and F. C. Zhang, Appl. Phys. Lett. 88, 162105 (2006).

\bibitem{Hamermesh} M. Hamermesh, Group theory and its application to
physical problems, (Reading, Mass. 1962).

\bibitem{Hankiewicz05prb} E. M. Hankiewicz, J. Li, T. Jungwirth, Q. Niu, S. Q. Shen, and J. Sinova, Phys. Rev. B 72,
155305 (2005).

\bibitem{KPFM} M. Nonnenmacher, M. P. Oboyle, and H. K. Wickramasinghe, Appl. Phys. Lett. 58, 2921
(1991); X. D. Cui, M. Freitag, R. Martel, L. Brus, and P. Avouris,
Nanoletter 3, 783 (2003).
\end{thebibliography}
\end{document}